\documentclass[preprint]{aastex}

\shorttitle{Candidate Li-rich Giants}
\shortauthors{Carbon, Gray, Nelson, and Henze}

\usepackage{natbib}
\begin{document}


\title{A Search for Candidate Li-Rich Giant Stars in SDSS DR10}



\author{Duane F. Carbon}
\affil{NASA Ames Research Center, NASA Advanced Supercomputing Facility, Moffett Field, CA, 94035-1000, USA;\\ Address correspondence to: Duane.F.Carbon@nasa.gov}
\author{Richard O. Gray} 
\affil{Appalachian State University, Dept. of Physics and Astronomy, CAP Bldg., Boone, NC 28608-0001;}
\author{Bron C. Nelson} 
\affil{NASA Ames Research Center, NASA Advanced Supercomputing Facility, Moffett Field, CA, 94035-1000, USA;}
\and 
\author{Christopher Henze}
\affil{NASA Ames Research Center, NASA Advanced Supercomputing Facility, Moffett Field, CA, 94035-1000, USA}


\begin{abstract}


We report the results of a search for candidate Li-rich giants among 569,738 stars of the SDSS DR10 dataset.  With small variations, our approach is based on that taken in an earlier search for EMP/CEMP stars and uses the same dataset.   As part of our investigation, we demonstrate a method for separating post-main sequence and main sequence stars cooler than $T_{\mathrm{eff}}$~$\approx$~5800~K~using our feature strength measures of the Sr~II~4078, Fe~I~4072, and Ca~I~4227 lines.  By taking carefully selected cuts in a multi-dimensional phase space, we isolate a sample of potential Li-rich giant stars.    From these, using detailed comparison with dwarf and giant MILES stars, and our own individual spectral classifications, we identify a set of high likelihood candidate Li-rich giant stars.  We offer these for further study to promote an understanding of these enigmatic objects.

\end{abstract}


\section{INTRODUCTION} \label{intro}

The many aspects of the evolution of the Li abundance from the Big
Bang to the present has generated a very large and complex literature.
The existence of Li-rich giants, first discovered by Wallerstein and
Sneden \citep{wallerstein82}, is one part of the Li story.  The origin
of these unexpected stars is still not fully understood.  Here we
outline the main points needed to appreciate the strangeness of
Li-enriched post-main sequence (post-MS) stars.  For those desiring a
deeper review, we recommend the useful and detailed summaries of the
observed and predicted post-MS evolution of the Li abundance given by
\citet{brown89, ruchti11, casey16} and the references cited therein.

The maximum Li abundances of F and G main sequence (MS) stars range from
A(Li)~$\approx$~2.1\footnote{A(Li)$~=~log_{10}(N_{Li}/N_{H})~+~12$  is the logarithmic abundance of Li on a scale where A(H)~=~12.0} for
ancient, low metallicity stars all the way up to A(Li)~$\approx$~3.3 for
young, metal rich stars \citep{lambert04,prantzos17}.  The Li
abundance is expected to change dramatically once stars leave the main
sequence.  Classical stellar evolution theory predicts that the Li
abundance will monotonically decrease once the bottom of the outer
convective envelope begins to move inward during post-MS
evolution.  As the envelope deepens, it increasingly entrains mass
zones where complete Li destruction has occurred.  This in turn
dilutes the MS surface abundance of Li.  By the time a star
reaches the base of the giant branch, the Li abundance is expected to
be only $\approx$~5-10\% of its MS value \citep{iben67a,iben67b}.  As
the star ascends the giant branch in the classical models, the convective envelope continues
to deepen and the Li abundances are predicted to drop even further.
Moreover, observational evidence \citep[e.g.,][]{brown89,mallik99,liu14}
indicates that the actual Li depletion is substantially greater than
the classical models predict for the majority of G and K giants.

With this background, it was a great surprise when Wallerstein and
Sneden in 1982 reported the discovery of a metal-rich, field K giant,
HD~112127, whose Li~I~6708 resonance doublet line had an equivalent
width of 0.45~\AA.  A model atmosphere abundance analysis of the
weaker Li~I~6104~transition led to A(Li)~$\approx$~3.0, very
substantially higher than expected theoretically.  \citet{kraft99}
reported the first Li-rich giant in a globular cluster, a star on the
first ascent of the giant branch in M3, with A(Li)~$\approx$~3.0 and a
Li~I~6708 equivalent width of 0.52~\AA.  Over the years, other Li-rich
giants have been found, both in Population I and Population II.  These
stars are quite rare however, comprising $\le$~1\% of giant stars
\citep{brown89,ruchti11,liu14,kirby16}.  The recent tabulation by
\citet{casey16} lists only 127 known giant stars with A(Li)~$>$~2.0;
the values of A(Li) in this listing range all the way up to
A(Li)~=~4.55.  The temperatures of these known Li-rich evolved
stars span a considerable range: the hottest is HD~172481, an F supergiant with
$T_{\mathrm{eff}}$~$\approx$~7250~K \citep{reyniers01} and the coolest
is IRAS~12556-7731, an M giant with $T_{\mathrm{eff}}$~$\approx$~3460~K
\citep{alcala11}.  Most known examples, however, are found among the G
and K stars where our search is optimized.

Numerous hypotheses have been put forward to explain these rare
Li-rich giants.  The majority appeal to some form of ``extra'' stellar
mixing on the giant or asymptotic giant branches that manages to
incorporate the Cameron-Fowler process \citep{cameron71} to generate
$^{7}$Li.  Other scenarios rely on acquiring Li-rich material from a
companion.  The reader is referred to \citet{kirby16} for a brief but
insightful review of all these mechanisms.

There is a need to find as many of these rare stars as possible to get
the full parameter space decribing their occurence and begin to narrow
down the numerous possibilities for producing them.  One of the most
successful efforts in finding new Li-rich giants was that undertaken
by Martell and Schetrone (M\&S13) \citep{martell13}.  They searched the
Sloan Digital Sky Survey Data Release 7 (SDSS DR7) \citep{abazajian09}
stars for candidates and identified 27 new Li-rich giants, 23 of which
they then subjected to high resolution abundance analysis.  It is the
purpose of the current paper to find additional Li-rich giant
candidates by carrying out a search using the data of SDSS DR10
\citep{ahn14}.  The DR10 dataset of optical stellar spectra is roughly
1.6X larger than that of the DR7 dataset.  This means that there are a
substantial number of additional stars now available to examine for
Li-enhancement.  We carry out our investigation using a slight
variation of the approach we employed earlier in a search for
extremely metal-poor (EMP) stars \citep{carbon17} (CHN17).

In Section~\ref{dataset}, we briefly describe how we processed the
DR10 dataset through our reduction pipeline so that we could extract
the individual feature measurements that are the basis of our
approach.  In the subsections of Section~\ref{extracting}, we
  detail how we chose our initial Li-rich candidates.  In particular,
  we describe how we selected (Section~\ref{luminosity}) and tested
  (Section~\ref{compMS13}) feature measurements to impose temperature
  and luminosity constraints, and how we extracted
  (Section~\ref{cuts}) a coarse sample of candidate Li-rich post-MS
  stars using a set of cuts in a multi-dimensional phase space.  Next we describe how we refined the sample to select
  only the most likely candidates  (Section~\ref{cull}) and then carried out a
  detailed spectral classification of these
  stars (Section~\ref{class}).  In Section~\ref{results} we discuss our final list of
  candidate post-MS Li-rich giants.  Our principal results are
  summarized in Section~\ref{summary}.

\section{THE DATASET} \label{dataset}

In our search for Li-rich giants, we use the previously prepared
dataset (CHN17) composed of calibrated optical fluxes and associated
data for 569,738 unique stellar spectra drawn from SDSS DR10.  The
associated data includes each star's coordinates, heliocentric
  radial velocity, median S/N, pixel-by-pixel inverse-variance values,
  and \emph{u,g,r,i,z} point spread function magnitudes (psfMag).  The reader is referred to Section
2 of CHN17 for details concerning the selection of the stellar data
from the whole SDSS DR10 database, the processing of the stellar
fluxes through our data reduction pipeline, and the feature strength
measurements that were subsequently made from the spectra.  Here, we
review briefly only the salient points needed for understanding the
arguments in the current paper.

The first step in our pipeline was establishing a continuum for
each spectrum.  Once the continuum level was established, it was
possible to compute quantitative measures for the spectral lines in
each spectrum.  Two types of feature measures were adopted.  The first was $S(\lambda_i)$
which is the fractional depth relative to the interpolated local
continuum of an individual spectral feature at wavelength $\lambda_i$.
The second was $D(\lambda_i)$ which is the depth of the line at
$\lambda_i$ in units of the local noise level in the spectrum as
determined from the spectrum's pixel-by-pixel inverse-variance values.
$S(\lambda_i)$ is a direct measure of the line's strength while
$D(\lambda_i)$ gives a handy measure for the line's strength
relative to the local noise level.  The latter can be particularly
helpful when dealing with intrinsically weak lines like the
Li~I~6708~line central to this paper.  Measurements of
$S(\lambda_i)$ and $D(\lambda_i)$ were made for 1659 spectral features
(each with a unique~$\lambda_i$) for all the 569,738 spectra of our
dataset.  This produced the final dataset of nearly 2~billion feature
measures used in this study.

In the CHN17 study, stars with specific interesting characteristics
were isolated from the above SDSS DR10 dataset by using linked scatter
plot (LSP) tools implemented on the NASA Advanced Supercomputing hyperwall.  (See CHN17, Section 1.1 for a detailed description of LSPs and the hyperwall.)  For example, by
making judiciously selected cuts in successive 2-D phase spaces, CHN17
were able to extract numerous candidate extremely metal poor (EMP)
stars from the general dataset.  Because of its flexibility, the LSP
method has powerful explorative capability.  For this reason, we used
LSPs on the hyperwall to make the initial reconnaissance of the
Li-rich giant problem.  It quickly became apparent that there were
indeed stars with strong Li~I~6708~lines in the DR10 dataset.
However, because the 569,738 spectra of the dataset include a very
wide range of temperatures, luminosities, and compositions, we needed
to determine how to extract candidate Li-rich giants from the rest of
the stars.  In the next section, we explain how we accomplished this.

\section{EXTRACTING THE CANDIDATE LI-RICH GIANTS} \label{extracting}

We require an approach which will effectively separate the relevant
post-MS stars from MS stars which may have comparable Li line
strengths.  The domain of chief interest is that occupied by the late
G and K giants.   We need to select feature
measures, $S(\lambda)$ and $D(\lambda)$, in our dataset that can be
used to isolate stars in this desired temperature and luminosity range.
Note that our approach relies solely on our feature measures.  We
specifically chose not to employ SDSS-provided quantities such as
$T_{\mathrm{eff}}$ and $\log~g$ simply because of the large errors that can
occur in individual values of these quantities (e.g., M\&S13, Appendix A).  The following
subsections detail how we used the feature measures to arrive at a
list of Li-rich giant candidates.

\subsection{MILES spectra to establish temperature and luminosity constraints} \label{luminosity}

\cite{gray09} note that the Ca~I~4227 line progressively strengthens
with decreasing temperature in going from G through K spectral types,
while the hydrogen lines progressively weaken over the same spectral
range.  This suggests that $S(\mathrm{Ca~I~4227})$ or $S(\mathrm{H~I})$ could be used as
a first-order surrogate for stellar temperature.  (We note here that
we investigated using colors based on the SDSS \emph{u,g,r,i,z}
magnitudes as temperature surrogates but found that they did not lead
to clean separation between MS and post-MS
stars.)  In order to estimate luminosity in G and K star domain,
\cite{gray09} and \citet{white07} suggest a number of possible metal
line strengths and ratios.  To determine whether any of these might be
helpful in our investigation, we turned to the MILES spectrum library
\citep{sanchez06} .

The 985 stars in the MILES library were selected for the purposes of stellar population synthesis.  As a result, they cover a wide range of temperatures, luminosities, and metallicities with particularly good coverage for the spectral ranges of most interest to us \citep[e.g.,][Figure 1]{sanchez06}.  The MILES spectra span the whole SDSS
optical wavelength range relevant to our study and have essentially
the same spectral resolution as the SDSS spectra.  Moreover,  the
stars in this library have carefully researched  $T_{\mathrm{eff}}$,
$\log~g$, and metallicity ([Fe/H]) drawn from the literature \citep{cenarro07}.  These attributes
make the MILES spectra ideal for determining which luminosity criteria
might be most effective for separating G and K MS stars
from post-MS stars.  To take advantage of the MILES
library, we ran the entire dataset of nearly 1000 MILES spectra through
the same spectral reduction pipeline that we used in our earlier
study.  The pipeline computed continua for each of the MILES spectra
and then computed $S(\lambda)$ and $D(\lambda)$ measures for each of
the 1659 spectral features we use.  Details of the pipeline process
may be found in CHN17, Sections 2.1-2.3.

To explore which of the Gray-Corbally and White et al. luminosity
criteria might be best for our purposes, we extracted two subsets from
our full MILES dataset of feature measurements.  The first subset, which we used to represent
MS stars, was comprised of the 344 MILES A through K stars with \citet{cenarro07}
$\log~g$~$>$~3.80.  The second subset, representing post-MS stars, was
comprised of the 254 MILES A through K stars with \citet{cenarro07} $\log~g$~$\le$~3.80.  The
division in $\log~g$ was chosen to be comparable to that adopted by
M\&S13 in isolating post-MS stars for their study.

\begin{figure}
\includegraphics[angle=-90,scale=.75]{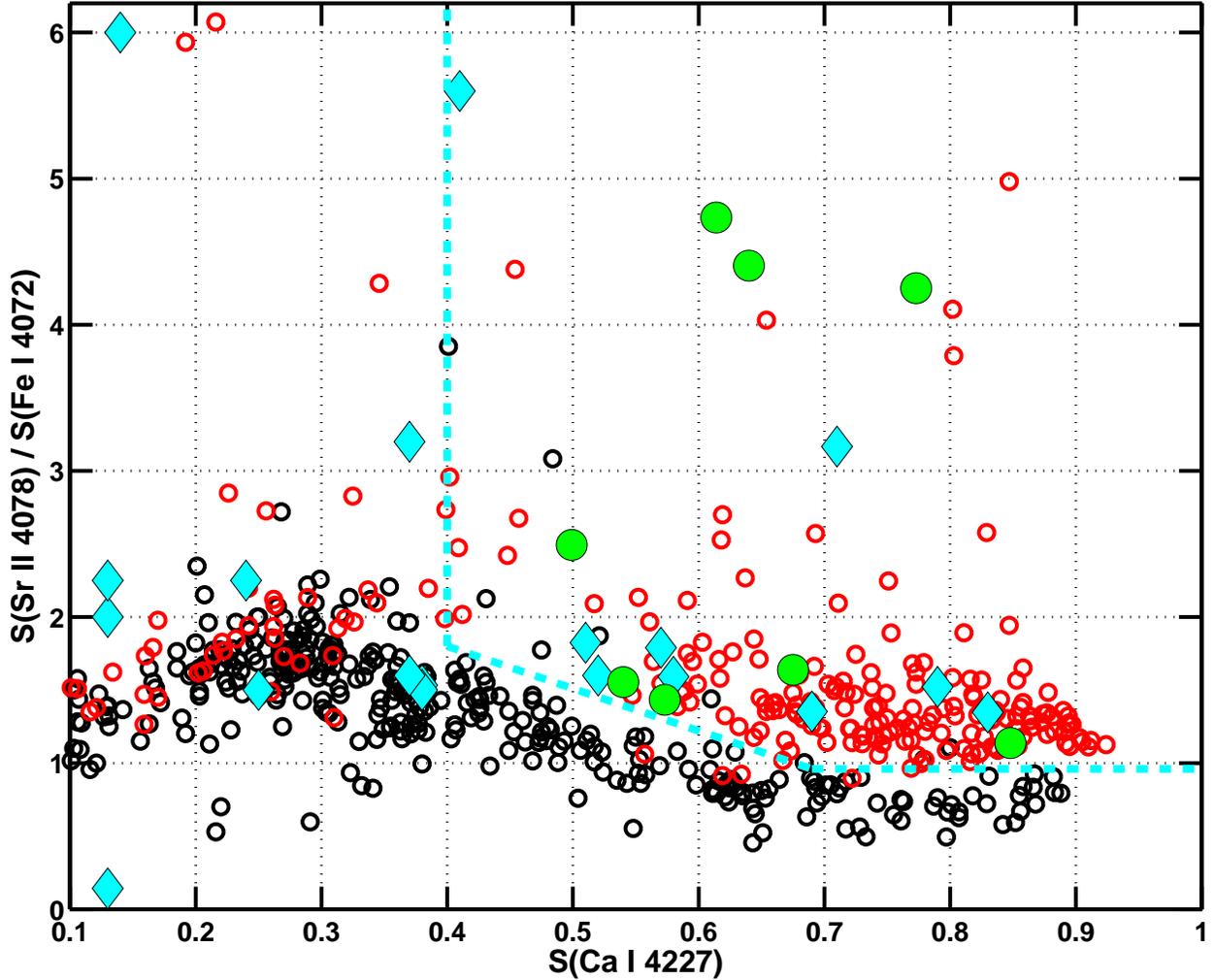}
\caption{Selected luminosity indicator,  $S(\mathrm{Sr~II~4078})/S(\mathrm{Fe~I~4072})$, versus $S(\mathrm{Ca~I~4227})$, a temperature surrogate.  The MILES  MS stars are plotted as black circles and the post-MS stars as red circles.   See text for details of how the MILES stars were divided according to $\log~g$.  The M\&S13 stars are plotted as cyan filled diamonds.  The region of this phase space where we searched for candidate Li-rich post-MS stars lies to the right of the vertical cyan dashed line at $S(\mathrm{Ca~I~4227})$~$=$~0.4 and above the lower bound defined by the cyan dashed lines running to the right.  The candidate Li-rich stars that we identify in Table~\ref{table_results} are displayed in this figure as green filled circles.  \label{luminfigure}}
\end{figure}

Many of the Gray-Corbally and White et al. luminosity criteria in the
G-K spectral range are ratios of lines strengths (or sums of line
strengths), e.g., the ratio of Y~II~4375 to Fe~I~4384.  We represented
these by taking the ratios of the corresponding line strength
measures, as in $S(\mathrm{Y~II~4375})/S(\mathrm{Fe~I~4384})$.  Using the MS
and post-MS subsets of MILES data, we examined the various
luminosity sensitive line ratios versus the feature strengths of the
likely temperature sensitive lines: $S(\mathrm{H{\alpha}})$, $S(\mathrm{H{\beta}})$,
$S(\mathrm{H{\gamma}})$, and $S(\mathrm{Ca~I~4227})$.  After considerable
experimentation, we found that $S(\mathrm{Sr~II~4078})/S(\mathrm{Fe~I~4072})$ vs
$S(\mathrm{Ca~I~4227})$ gave the clearest separation between MS and
post-MS stars for the G-K stars.  We show this separation in
Figure~\ref{luminfigure}.   The luminosity sensitivity of the $S(\mathrm{Sr~II~4078})/S(\mathrm{Fe~I~4072})$ ratio is a result of the rather different electron pressure sensitivities of these two lines in the cooler stars. The line ratio systematically shifts as the gravity, and hence electron pressure, decreases with increasing luminosity.  A discussion of such effects may be found in \citet{gray92}, for example.

The locations of the MS and post-MS stars (black
and red circles, respectively) in Figure~\ref{luminfigure}, show
that, while there are a few exceptions, the Sr~II/Fe~I ratio nicely
discriminates between MS and post-MS stars in
the region $S(\mathrm{Ca~I~4227})$~$\ge$~0.4.  In contrast, the Sr II/Fe I ratio
becomes unreliable as a luminosity diagnostic for
$S(\mathrm{Ca~I~4227})$~$<$~0.4.  Examination of the $S(\mathrm{Ca~I~4227})$ vs $T_{\mathrm{eff}}$
relation for the MILES stars indicates that the $S(\mathrm{Ca~I~4227})$~=~0.4
boundary occurs at $\approx$~5800~K, a temperature that
corresponds to early-mid G stars in the case of dwarfs
\citep{boyajian12}.  Thus, since it places us in the stellar
temperature range most relevant to our search, restricting our search
to stars with $S(\mathrm{Ca~I~4227})$~$\ge$~0.4 should pose no difficulty.
However, in the next subsection, we will note one important caveat.

\subsection{Comparison with confirmed Li-rich giants}\label{compMS13}

It is helpful to illustrate how a known set of Li-rich giants are
distributed in Figure~\ref{luminfigure}.  M\&S13 searched for
Li-rich giants among the stars of SDSS DR7.  They chose a set of 8535 stars from
DR7 whose SEGUE Stellar Parameter Pipepine (SSPP) $T_{\mathrm{eff}}$ and
$\log~g$ values indicated that they would be red giant branch (RGB)
stars lying somewhere between slightly below the red giant bump and
the red giant tip.  They estimated the Li~6708~strength in these
stars using a spectral index they computed centered on the line.
Selecting only those stars with the most promising Li spectral indices
(162 stars), they used low resolution spectrum synthesis to sub-select
a set of 36 for follow-up high resolution study.  Of these 36, they
confirmed that 27 were indeed Li-rich based on high resolution
spectroscopy and spectrum synthesis.  (We note that M\&S13 present
derived Li abundances for only 23 of the 27 stars because of S/N
problems.  Nevertheless, we will consider all 27 as ``confirmed Li
rich'' as indicated in their Table 1.)  Of these 27 SDSS DR7 Li-rich stars, 19 are
included in the download for our DR10 dataset, the 8 missing stars
violated one or more of the selection criteria we adopted in selecting
the stars for our dataset.

 In Figure~\ref{luminfigure} we show the 19 M\&S13 stars as cyan
 diamonds.  We see that 9 of the 19 M\&S13 stars fall comfortably in
 the region occupied by post-MS stars with $S(\mathrm{Ca~I~4227})$~$\ge$~0.4.
 The remainder, with smaller $S(\mathrm{Ca~I~4227})$ values, fall in the region
 where the $S(\mathrm{Sr~II~4078})/S(\mathrm{Fe~I~4072})$ ratio does not reliably
 distinguish luminosities.  Eight of the 10 stars in this latter
 region have [Fe/H] determined by M\&S13 (their Table 2).  Six
 have [Fe/H] ranging from -1.4 to -2.6, i.e., they are metal poor.
 The other 2 stars have the highest $T_{\mathrm{eff}}$ values (5250~K, 5625~K) of the stars with
 $S(\mathrm{Ca~I~4227})$~$<$~0.4 as well as depressed [Fe/H] ($\le$~-0.29).
 (The 2 stars that do not have M\&S13 [Fe/H] values, have SDSS
 [Fe/H] of -0.66 and -1.50.)  Both low metallicity and higher
 $T_{\mathrm{eff}}$ could explain the presence of M\&S13 stars to the left of
 the $S(\mathrm{Ca~I~4227})$~$=$~0.4 boundary.  All of the stars to the right
 of the $S(\mathrm{Ca~I~4227})$~$=$~0.4 boundary have [Fe/H], as determined
 by M\&S13, ranging from -0.47 to +0.41.  This suggests that, by
 restricting ourselves to stars with $S(\mathrm{Ca~I~4227})$~$\ge$~0.4, we may run
 the risk of missing Li-rich giants with low metallicities.  Since we
 see no obvious way at this time to devise a luminosity criterion that
 does not risk excluding such stars, we shall proceed.  Low
 metallicity giants that are sufficiently cool (hence having
 intrinsically stronger Ca~I~4227) might still land to the right of
 the $S(\mathrm{Ca~I~4227})$~$=$~0.4 boundary and thus be detectable by us.

\subsection{Imposing the final constraints} \label{cuts}

In order to arrive at a useful set of Li-rich giant candidates it is
necessary to constrain more than just the value of $S(\mathrm{Li~I~6708})$ and
the region in $S(\mathrm{Sr~II~4078})/S(\mathrm{Fe~I~4072})$ vs $S(\mathrm{Ca~I~4227})$ space.
For good quality results, one must also add constraints on the noise
levels both overall and locally in the 6700~\AA~and 4070~\AA~regions.
Similarly, it is necessary to guard against TiO contamination of the
Li~I~6708 line region.  After considerable experimentation, we chose
the set of constraints listed below.  The additional constraints
eliminated many spectra in which noise/contamination produced uncertain results
for the value of $S(\mathrm{Li~I~6708})$ and/or the
$S(\mathrm{Sr~II~4078})/S(\mathrm{Fe~I~4072})$ ratio.

The measures we selected and the constraints we imposed on them are
summarized in the logical expressions below. All of these constraints,
1~-~6, are applied at the same time to the feature measures of the
569,738 stars in the dataset.  Only stars that simultaneously satisfy
all the specified constraints are considered in the remaining discussion.


\begin{mathletters}
\begin{equation}
     ( 0.4 \le S(\mathrm{Ca~I~4227}) < 0.69 )  ~\&~ ((S(\mathrm{Sr~II~4078})/S(\mathrm{Fe~I~4072}))  \ge (2.96 - 2.90 * S(\mathrm{Ca~I~4227}) ) )
\end{equation}
\begin{equation}
    (S(\mathrm{Ca~I~4227}) \ge 0.69) ~\&~ (0.96 < (S(\mathrm{Sr~II~4078}) / S(\mathrm{Fe~I~4072}) ) ) 
\end{equation}
\end{mathletters}
\begin{equation}
    S(\mathrm{Li~I~6708}) \ge 0.04
\end{equation}
\begin{equation}
   \mathrm{median~(S/N)}   > 20   
   \label{medianSN}
\end{equation}
\begin{equation}
     D(\mathrm{Li~I~6708}) > 1.0
\end{equation}
\begin{equation}
    S(S\mathrm{r~II~4078}) > 0.0 
\end{equation}
\begin{equation}
   S(\mathrm{TiO~6815}) < S(\mathrm{Li~I~6708})
\end{equation}

We now briefly describe the rationale for each cut shown above:

Constraints 1a and 1b apply the luminosity discriminant ratio
$S(\mathrm{Sr~II~4078})$/$S(\mathrm{Fe~I~4072})$  described in Section \ref{luminosity}.  Constraint
1a applies to the left-hand portion of the region outlined by cyan
dashed lines in Figure~\ref{luminfigure} (i.e., 0.4~$\le$~$S(\mathrm{Ca~I~4227})$~$<$~0.69 and above the sloping cyan dashed line); 1b applies to the right-hand
portion (i.e., $S(\mathrm{Ca~I~4227})$~$\ge$~0.69 and above the horizontal cyan dashed line).

Constraint 2 further selects out those stars which have Li I
absorption strengths above a minimum threshold.  We selected
  the threshold to be equal to the $S(\mathrm{Li~I~6708})$ measure of
  the M\&S13 Li-rich giant with the weakest Li~I feature.

Constraint 3 isolates the objects which have sufficient S/N in their
spectra to make estimating luminosity and Li~I strength more robust.
We found that the spectra of stars with poorer S/N are generally much too noisy to
yield reliable measures of either $S(\mathrm{Sr~II})$/$S(\mathrm{Fe~I})$ or $S(\mathrm{Li~I})$.

Constraint 4 further limits the subset of stars to those with Li~I~6708
line depths more than 1~$\sigma$ above the local noise level.  This helps
eliminate stars that have excessive noise near the Li~I line.

Constraint 5 limits the stars to only those with detected Sr~II~4078 absorption, removing stars for which random noise, or poor continuum placement, produces a false emission feature.

Constraint 6, which uses a $^{48}$Ti$^{16}$O (3,2) gamma system band head, was introduced to bias against stars for which the TiO bands were becoming sufficiently strong that they were noticeably affecting the region of the Li I line.  

To impose the constraints described above, we constructed a suite of
MATLAB\copyright~\citep{matlab11} codes implemented on a single
computer workstation.  The suite reads in the constraints on specified
feature variables and returns a list of those stars for which the
specified variables simultaneously satisfy all the constraints.  This
is logically equivalent to the CHN17 method of making a series of
successive cuts in 2-D phase spaces that was the basis of the LSP
approach.  Applying the constraints 1 through 6 to the dataset of
569,738 SDSS DR10 stars produces a subset of 1,523 stars which are
potentially Li-rich giants.  In the next sub-section, we will describe
how we select out the most likely candidate Li-rich giants.

\subsection{Extracting the Candidate Li--rich Giants} \label{cull}

\subsubsection{Eliminating the obvious false positives} \label{coarse_cull}


The feature constraint on median S/N in Equation~(\ref{medianSN})
was intentionally left ``softer'' than it might have been so as to
capture as many candidates as possible.  However, this means that
stars may slip through the constraints whose spectra are too
contaminated by noise in crucial spectral regions to be sure of their
status.  In addition, some of the feature strengths used in the
constraints may have erroneous values caused by poor continuum
placement.  (A more detailed discussion of these issues may be found
in CHN17, Section 4.)  We dealt with these issues by visually examining
the spectra of each of the 1,523 stars selected by the constraints of
the previous section.  The visual examination was done in two
steps. 

In the first step, the chief criteria were the strength and
apparent position of the purported Li I line, whether the spectral
regions around Li I line and the luminosity indicators appeared
relatively unaffected by noise, whether there appeared to be TiO
contamination of the Li region, and whether the continuum placement
was appropriate.  A secondary consideration was whether the
Li~I~6708~line was comparable to or stronger than Ca~I~6718~line (see
\citet[][Figure 5]{casey16}).  The ratio of these two lines was used
by \citet{kumar11} in their study to identify candidate Li-rich stars
from low-resolution giant spectra. This coarse initial cull was straightforward and was accomplished relatively quickly.  It
eliminated 1,350 stars from further consideration, the vast majority because
the local noise level was too large to be confident of the Li line
strength.

In the second step, the spectra of the 173 remaining stars were
subjected to a more prolonged and careful visual inspection which
concentrated on the position, shape, and strength of their Li~I 6708
line and the quality of the spectrum.  Stars were eliminated if the
apparent Li~I feature appeared to be strongly asymmetric, shifted
significantly from its nominal position, or was too similar in appearance
to the surrounding noise features.  This second visual cull left 49
candidates.  These 49 stars included \emph{all} 9 of the M\&S13
Li-rich giants which fell into our search region, evidence that our
selection procedure was working well.  The next sub-sections describe
how we confirmed whether the final 40 previously unrecognized Li-rich
candidates were indeed giants.


\subsubsection{Comparison with MILES stars} \label{comp_miles}

To increase our confidence in the likelihood that we were selecting
stars that were good Li-rich giant candidates, we first carried
out a systematic comparison of each of the 40 stars with the MILES MS
and post-MS stars.   First we normalized the spectrum of each Li-rich giant candidate and each MILES MS and post-MS star.  This was accomplished by normalizing each spectrum by its continuum and then by its flux at 5837 \AA\  so as to keep
the scales of the different spectra consistent.  Next we interpolated the resulting MILES spectra onto the SDSS DR10 wavelength set over the interval [3850~-~7400~\AA].  Using the resulting fluxes, we computed the
following summed square differences, SSD, for each of the 40 stars
against each of the spectra of all the MILES MS stars seriatim and
then, separately, all the MILES post-MS stars:
\begin{displaymath}
\mathrm{SSD} = \sum_{k=1}^n [F_{\mathrm{cand}}(\lambda_k) - F_{\mathrm{MILES}}(\lambda_k)]^2,
\end{displaymath}
where $F_{\mathrm{cand}}(\lambda_k)$ is the normalized flux at wavelength
$\lambda_k$ of one of the candidate Li-rich stars, $F_{\mathrm{MILES}}(\lambda_k)$
is the corresponding normalized flux of one of the MILES stars, and n
is the number of wavelengths in the common wavelength set.

For each candidate Li-rich star, we compared its spectrum with the
closest matching (i.e., smallest SSD values) MILES MS and post-MS
stars to see whether the candidate spectrum appeared more consistent
with the spectra of dwarfs or giants.  Attention was paid not only to
the Sr~II~4078/Fe~I~4072 ratio, but also to the strengths of
Sr~II~4078 relative to the Fe~I~4064 and Fe~I~4046 lines
\citep{gray09}.  We also considered the values of SSD for the
candidate star and the ten closest matches from the MILES main
sequence and post-MS lists.  For many stars, the SSD values for the
top ten closest matches were very strongly in favor of a candidate
being most like a giant or dwarf.

Based on the above comparisons, 31 stars were rejected because
  they more closely matched the spectra of MILES MS stars both in
  their Sr~II to Fe~I line ratios and in their SSD values. Nine stars
  remained as candidates to be Li-rich giants.  We decided it was
  prudent to subject these 9 stars to an additional final check.  The
  next sub-section describes the effort by one of us (ROG) to examine
  the spectra of the 9 stars in detail and make definitive spectral
type classifications based on more than the limited number of spectral
features we have considered up to this point.

\subsection{Detailed spectral classification} \label{class}

While the SDSS spectra have a much larger spectral range,
the most sensitive temperature and luminosity criteria are found in the
violet -- green region, 3800 -- 5600\AA.  Because of the unavailability of
an MK standard star library for the SDSS spectra, we convolved the SDSS 
spectra with a gaussian to reduce the resolution to that of the {\tt libnor36}
MK Standards library  \footnote{That library, as well as other MK standards 
libraries may be downloaded as ascii files ({\tt mklib.tar.gz}) from
\url{http://www/appstate.edu/$\sim$grayro/mkclass}.  The MK standards used in {\tt libnor36} are listed on that
same site.} (3.6\AA/2 pixels) of \citet{gray14}.

\citet{gray09} detail the temperature and luminosity criteria used
in the MK classification of G- and K-type stars.  In summary, temperature
criteria involve the ratio of low-excitation neutral metal lines to hydrogen
lines (\ion{Fe}{1} $\lambda 4046$/H$\delta$, 
\ion{Fe}{1} $\lambda 4144$/H$\delta$, \ion{Fe}{1} $\lambda 4383$/H$\gamma$
as well as similar line ratios in the vicinity of H$\beta$).  Those ratios, 
however, are invalid in metal-poor stars, and in that case, the ratio of
lines of the \ion{Cr}{1} triplet ($\lambda\lambda$4254, 4275, 4290 -- all
resonance transitions) with the higher-excitation 
\ion{Fe}{1} $\lambda\lambda 4250$, 4260, and 4326 lines provide 
metallicity-independent temperature criteria.  Luminosity criteria include 
the ratios of 
\ion{Sr}{2} $\lambda 4077$\footnote{Note that the wavelengths used in this sub-section are those adopted by \citet{gray09} and may differ slightly from the air wavelengths used elsewhere in the paper which are rounded to the nearest \AA.} to nearby \ion{Fe}{1} lines ($\lambda\lambda 4046$,
4064, and 4072), \ion{Sr}{2} $\lambda 4216$/\ion{Ca}{1} $\lambda 4226$, 
\ion{Y}{2} $\lambda 4376$/\ion{Fe}{1} $\lambda 4383$ as well as the strength of
CN violet system, in particular the band blueward of the $\lambda 4215$ 
bandhead.  However, in stars with carbon abundance peculiarities, the CN 
band strength can give spurious results, as proved to be the case with a 
number of stars in the candidate Li-rich sample under consideration.

The spectral types were determined by eye on the computer screen by
direct comparison with the {\tt libnor36} MK standards.  The spectral
types we obtained are listed in Table 1 for the 8 candidates which
proved to be giants.  One candidate (J215914.37+004515.8) turned out
to be a G9 dwarf and will not be considered further.  Three out of the
final 8 appear to be normal late G- and early K-type giants.  The
remaining stars, all late G- to early K-type giants (with the
exception of J150029.54+010744.8, which is a Ib-II supergiant), show
carbon peculiarities in the form of weak CH (G-band) and CN bands.

\begin{deluxetable}{ccrrrrllc}
\tabletypesize{\scriptsize}
\tablecaption{Candidate Li-Rich Post-MS Stars \label{table_results}}
\tablehead{\colhead{} & \colhead{Obs} & \colhead{$S$} & \colhead{$D$} &  \colhead{$S$} & \colhead{$\mathrm{Sr~II}$} &  \colhead{SDSS} & \colhead{Our} & \colhead{SpT} \\
\colhead{Star Name} & \colhead{Date} & \colhead{$\mathrm{Li~I}$} & \colhead{$\mathrm{LI~I}$} &  \colhead{$\mathrm{Ca~I}$} & \colhead{$\mathrm{/Fe~I}$} &  \colhead{$\mathrm{SpT}$} & \colhead{$\mathrm{SpT}$} & \colhead{Notes}\\
}



\startdata 

J021646.38$-$003333.5   & 10-10-2010   & 0.14  & 10.7  & 0.68  &  1.64  &  K4III  &  K0 III-IV  & \nodata \\
J060724.43$+$240052.4  & 02-25-2008   & 0.08  &  6.7   & 0.54  &  1.55  &  K1  &  G9 III-IV  & \nodata \\
J062219.56$+$414403.8  & 12-07-2007  & 0.05  & 3.6   &  0.50  &  2.49  &  K1  &  G8 II-III  & \nodata \\
J092210.66$+$162455.9  & 03-12-2012  & 0.11  & 6.6   &  0.64  &  4.40  &  K4III  &  G9 II-III  CN-1 & \tablenotemark{1} \\ 
J122728.00$+$054420.2  & 04-20-2009  & 0.14  & 10.6  & 0.85  &  1.14  &  K5  &  K0 III CN-1 CH-1 &  \tablenotemark{2} \\ 
J143237.11$+$024533.4  & 03-31-2001  & 0.15  & 8.2   &  0.77  &  4.25  &  K5  &  K0 II-III CN-2 CH-2 & \tablenotemark{3} \\ 
J150029.54$+$010744.8  & 05-19-2009  & 0.17  & 12.7  & 0.61  &  4.73  &  K5   &  G8 Ib-II CN-1 & \tablenotemark{4} \\  
J183259.15$+$222243.5  & 07-01-2006  & 0.11  & 4.1   &  0.57  &  1.43  &  K1  &  G8 III CN-0.5 &  \tablenotemark{5} \\

\enddata
\tablenotetext{1} {CN band weak.}
\tablenotetext{2} {Both CN and CH bands weak.}
\tablenotetext{3}{Both CN and CH bands markedly weak.}
\tablenotetext{4}{CN band weak for luminosity type.}
\tablenotetext{5} {CN band slightly weak.}
\end{deluxetable}

\section{RESULTS}\label{results}

\begin{figure}
\includegraphics[angle=-90,scale=.50]{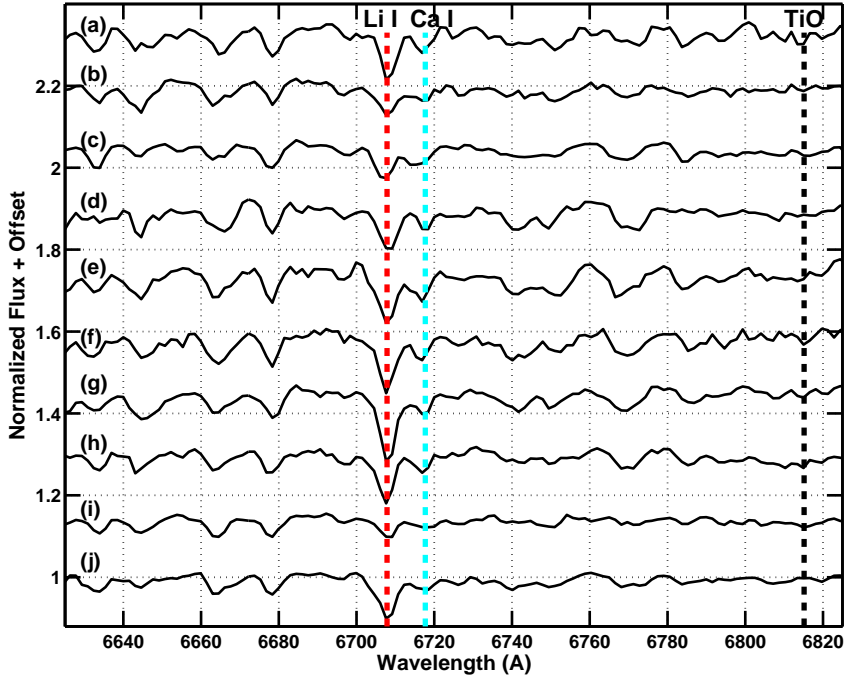}
\caption{ Comparison of the stars in Table~\ref{table_results} with two bounding M\&S13 stars.  The spectra have been normalized to their mean flux in the interval [6693,6695].  The Table~\ref{table_results} stars are: (a) J021646.38-003333.5; (b) J060724.43+240052.4; (c) J062219.56+414403.8; (d) J092210.66+162455.9; (e) J122728.00+054420.2; (f) J143237.11+024533.4; (g) J150029.54+010744.8; (h) J183259.15+222243.5. The bottom two stars are the M\&S13 high S/N stars with the weakest and strongest Li~I~6708 lines: (i) J030437.40+382346.1, and (j) J051523.18+155855.4, respectively. \label{spectra}}
\end{figure}

The final set of 8 stars that survived the vetting process described
in the previous section are presented in Table~\ref{table_results}.
For completeness, we retain  J150029.54+010744.8 in the set of candidates despite its luminosity class.  A model atmosphere analysis will be needed to accurately place it relative to the giant branch.  The table gives the date of observation of the measured SDSS DR10
spectrum, selected feature strengths and ratios as described in the
text, the SDSS-assigned spectral type, and the spectral type
determined by us.  The $S(\mathrm{Li~I~6708})$ values show that, despite the
comparatively low resolution of the SDSS spectra, the absorption
depths of the Li I lines in the candidates are not trivial, ranging
from ~5\% to 17\%.  The $D(\mathrm{Li~I~6708})$ values, the line depth in
units of the local noise level, all suggest solid detections.   Comparing the
two columns of spectral types in the table, it is immediately apparent that our
spectral types are all systematically earlier than the SDSS
assignments.  The differences are generally small and perhaps
partially reflect the coarseness of the ELODIE library used by SDSS to
classify the stars \citep{lee08}.  Nuances introduced by weakening of
CH and CN bands within a spectral type, captured by our approach and
indicated in the ``SpT Notes'' column, might have confused the SDSS
classification as well.

The DR7 dataset used by M\&S13 contains observations obtained up to
July 2008 \citep{abazajian09}, whereas the DR10 dataset we used
contains SDSS optical observations through June 2012 \citep{ahn14}.
As we mentioned in Section~\ref{coarse_cull}, our approach captures
all 9 M\&S13 stars in our dataset which have
$S(\mathrm{Ca~I~4227})~\ge~0.4$.  We note that
Table~\ref{table_results} contains 4 additional stars that, according
to their dates of observation, were present in the DR7 dataset.  These
stars apparently failed to pass one of the selection criteria used by
M\&S13 to derive their list of 36 Li-rich candidates suitable for high
resolution examination.  It will be interesting to see whether or not
future analysis of these stars confirms that they are Li-rich giants
as we suggest.  The spectra of the remaining 4 stars in
Table~\ref{table_results} were obtained after July 2008 and could not
have been considered by M\&S13.  We find it somewhat surprising that
we discovered only 4 new candidates among the stars observed after the
end of the DR7 dataset.  Our downloaded CHN17 dataset has 364,265
stars observed before July 2008 and 205,473 stars observed after that
date.  This makes the post-July 2008 portion of the dataset 56\% of
the size of the earlier portion.  Given that we found a total of 13
Li-rich candidates in the earlier dataset (the 9 M\&S13 stars plus our
4 new candidates), one naively might expect that the more recent
portion alone would yield roughly 7 candidate Li-rich giants.  That we
found only 4 may be only a reflection of the uncertainty of small
number statistics.  It also may be the result of a shift in the
spectral type mix between the two portions of the dataset given that
the stellar classes targeted by the SDSS changed with time as the
survey went on.

We show in Figure~\ref{spectra} spectra of the stars of
Table~\ref{table_results} in the vicinity of the Li I line.  We have
marked with dashed lines the Li~I~6708 doublet, the Ca~I~6718 feature
used by \citet{kumar11} and used by us as a secondary criterion, and
the TiO~6815 band head we used in Section~\ref{cuts}.  For comparison,
we also show at the bottom the two high S/N M\&S13 stars with
$S(\mathrm{Ca~I~4227})$~$\ge$~0.4 having the weakest and
the strongest Li I lines.  It is apparent that
the Li I features in our candidates are comparable in strength or
stronger than those in stars identified as Li-rich giants by M\&S13.

Finally, we show our Li-rich giant candidates in
Figure~\ref{luminfigure} as green dots.  Our 8 candidate Li-rich
giants are distributed in the plot much like the 9 M\&S13
already-confirmed Li-rich giants.  The most luminous are well away
from the boundary between the MS and the post-MS stars.  Like the
majority of M\&S13 stars, the remainder of our candidates lie
closer to the boundary.  The locations of our candidates in
Figure~\ref{luminfigure}, their spectral types in
Table~\ref{table_results}, and their strong Li I lines
(Figure~\ref{spectra}) all suggest that they are Li-rich giants.  We
offer these candidates to researchers for closer examination, an
undertaking well beyond the limited scope of this paper.  Model
atmosphere analyses of higher resolution spectra will be required to
definitely determine the Li abundance and evolutionary status of our
candidate Li-rich giants.

\section{SUMMARY} \label{summary}

In the current paper, we describe a new approach to identifying
candidate Li-rich giants using the SDSS DR10 data release.  As part of an
earlier investigation (CHN17), 569,738 SDSS DR10 spectra were
processed through a pipeline which yielded feature strength
measurements for each of 1659 unique spectral features in each
spectrum.  The resulting nearly 2~billion feature measurements can be
used to construct phase spaces of measurements.  One may then
introduce constraints that can be used to isolate stars with desired
characteristics.  In CHN17 this was accomplished using
linked scatter plots and the hyperwall.  In the current paper, we
introduced a simple procedure for applying constraints that can be
accomplished on a single workstation.

Guided by the literature on spectral classification of low resolution
spectra, we searched for feature measurements that could be used to
identify G and K giants, the subset of stars expected to harbor the
Li-rich giants.  Using the MILES spectra for MS and post-MS
stars,  we identified the $S(\mathrm{Ca~I~4227})$ feature strength as the best
surrogate for $T_{\mathrm{eff}}$.  After considerable experimentation, we found
that the $S(\mathrm{Sr~II~4078})/S(\mathrm{Fe~I~4072})$ ratio gave reasonable separation
between MS and post-MS stars for stars with
$T_{\mathrm{eff}}$~$\lesssim$~5800~K.  Armed with this insight, we isolated 1,523
potential Li-rich giants by applying constraints using only the
feature strengths of Li~I~6708, Ca~I~4227, Sr~II~4078, Fe~I~4072, and
TiO~6815, as well as the median and local S/N values.

Visual inspection of the spectra quickly reduced the list of possible
candidates to 49 stars, 9 of which were Li-rich giants already
discovered by M\&S13.  Next, the remaining 40 stars were systematically
compared with those MILES dwarf and giant stars most similar to them
in overall spectral energy distribution.  Stars selected as most
compatible with giants were then carefully classified for spectral
type.  These steps produced the 8 candidate Li-rich giants shown in
Table~\ref{table_results}.  We strongly recommend that researchers
interested in expanding the list of known Li-rich giants consider
these stars for detailed high resolution investigation.

\vspace{5mm}

\acknowledgments

The authors wish to thank Karen A. Huyser, Winifred M. Huo, and David
W. Schwenke for their insightful comments on the draft of this paper.
The authors also wish to thank the anonymous referee for numerous
thoughtful suggestions that significantly improved the text.  Funding for SDSS-III
has been provided by the Alfred P. Sloan Foundation, the Participating
Institutions, the National Science Foundation, and the U.S. Department
of Energy Office of Science. The SDSS-III web site is
http://www.sdss3.org/.  SDSS-III is managed by the Astrophysical
Research Consortium for the Participating Institutions of the SDSS-III
Collaboration including the University of Arizona, the Brazilian
Participation Group, Brookhaven National Laboratory, Carnegie Mellon
University, University of Florida, the French Participation Group, the
German Participation Group, Harvard University, the Instituto de
Astrofisica de Canarias, the Michigan State/Notre Dame/JINA
Participation Group, Johns Hopkins University, Lawrence Berkeley
National Laboratory, Max Planck Institute for Astrophysics, Max Planck
Institute for Extraterrestrial Physics, New Mexico State University,
New York University, Ohio State University, Pennsylvania State
University, University of Portsmouth, Princeton University, the
Spanish Participation Group, University of Tokyo, University of Utah,
Vanderbilt University, University of Virginia, University of
Washington, and Yale University.  This research has made use of the
SIMBAD database, operated at CDS, Strasbourg, France.
MATLAB\copyright~2015 The MathWorks, Inc. MATLAB and Simulink are
registered trademarks of The MathWorks, Inc. See
www.mathworks.com/trademarks for a list of additional
trademarks. Other product or brand names may be trademarks or
registered trademarks of their respective holders.

\vspace{5mm}



{\itshape Facilities:} \facility{Sloan}





\bibliography{li_rich_paper_revised_arxiv_version}


\clearpage

\end{document}